\documentclass[aps,prd,twocolumn,floats,floatfix,tightenlines]{revtex4}
 
\usepackage{graphics}
\usepackage{epsfig}
\usepackage{amssymb}
\usepackage{mathrsfs}
\usepackage{dsfont}
\usepackage{wasysym}
\usepackage{latexsym}
\input epsf
\usepackage{epstopdf}
 
\newcommand{\II}{\mathrm{I\!I}}

\begin{document}

\title{Observer dependence of the quasi-local energy and momentum in
  Schwarzschild space-time} 
\author{P.~P. Yu\footnote{ppyu@dartmouth.edu}}
\author{R.~R. Caldwell}
\affiliation{Dartmouth College, Hanover, NH 03755 USA}

\date{\today}

\begin{abstract}

The observer dependence of the quasi-local energy (QLE) and momentum in the
Schwarzschild geometry is illustrated. Using the Brown-York prescription,
the QLE for families of non-geodesic and geodesic observers penetrating the
event horizon is obtained. An explicit shell-building process is
presented and the binding energy is computed in terms of the QLE
measured by a static observer field at a radius outside the horizon
radius. The QLE for a radially geodesic observer field freely-falling
from infinity is shown to vanish. Finally, a simple relation
for the dynamics of the quasi-local momentum density for a geodesic
observer field is noted. 

\end{abstract}



\maketitle

\section{Introduction}

There has been rising interest in studying the quasi-local energy (QLE)
inside the event horizon of the Schwarzschild geometry
\cite{Lundgren:2006fu,Blau:2007wj}. However, the absence of physical, static
observers on the interior, as exist on the exterior, has prevented a simple
passage of QLE across the horizon. The alternative is then to identify
families of non-static, non-geodesic and geodesic observers that can
actually penetrate the horizon smoothly and measure the QLE. This is the
main focus of the present work. It is hoped that this note can bring to the
ongoing discussion a somewhat different perspective.

It is well-known that the quasi-local energy is observer dependent
\cite{Brown:1992br}. Take the Brown-York QLE for example, a family of
prescribed observers -- interchangeably referred to as an observer field --
on a closed, orientable, space-like 2-surface, $S$, as embedded in a
space-like hypersurface $\Sigma$ in the space-time manifold $M$, set
up their frame fields and evaluate the trace of the mean curvature of
$S$ as embedded in $\Sigma$ at their respective locations. After
averaging over $S$ and calibrating against the Minkowski space
reference, the ultimate quantity is the measured QLE
associated with $S$. Among a sea of available observers in a
space-time, the most physically meaningful ones are the observer
fields that are static and those that are geodesic. In the
Schwarzschild space-time, no static observer fields can enter the
event horizon; on the other hand, there do exist geodesic observer
fields that can cross the event horizon. Hence, to use one observer
field as a probe to measure the QLE in the
Schwarzschild space-time, the best candidates would be a non-static,
observer field that penetrates the horizon or a geodesic observer
field.  

The Brown-York QLE measured by the above two kinds of
observer fields in the Schwarzschild geometry are examined in this article.
Generalizations to other spherically symmetric space-times should be
straightforward. A simple observation regarding the dynamics of the 
relative quasi-local momentum density in a geodesic observer field is also
presented.

A note about the Liu-Yau QLE \cite{Liu:2003} in the
Schwarzschild space-time is recorded here as a comparison. Since the
construction of the physical part of the Liu-Yau QLE employs a
co-dimension 2 embedding of $S$, it is independent, in the exterior of
the horizon, of the choice of observer field. However, the Liu-Yau
QLE becomes invalid in the interior since the mean curvature vector of
$S$ in $M$ becomes time-like, violating one of its defining
hypotheses.  

The structure of this article is as follows. An expeditious review of the
fundamentals of the Schwarzschild and Kruskal space-times is given in
Sec.~\ref{review}. Static and non-static, non-geodesic observer fields are
used in the study in Sec.~\ref{nongeodesic}, whereas the case of a geodesic
observer field is presented in Sec.~\ref{geodesic}. Physical interpretations
of the results are summarized in Sec.~\ref{summary}.

\section{A Review of Definitions}
\label{review}

Collected here are some necessary definitions and facts that are frequently
referred to throughout the article. Hence, notations and terminology are
unified in a consistent fashion, at the outset. It should be duly 
acknowledged that equivalent co-ordinate systems, such as those due to
Novikov, Painleve, and Eddington-Finklestein, may as well be adopted
for the problem at hand. The ones employed in the present work are an
impartial choice at the service of highlighting the geometric 
nature. Details can be found in Ref.~\cite{SachsWu1977}, for example.  

The Schwarzschild space-time $({\cal M}, g, D)$ consists of two connected
components ${\cal M}=N \cup B$, namely its exterior $N=P_{\rm I}\times_{\rm
  r} S^2$, where $P_{\rm   I}=p^{-1}({\mathds R}^1\times (2{\rm M},
+\infty))$, and its interior $B=P_{\rm II}\times_{\rm r} S^2$, where
$P_{\rm II}=p^{-1}({\mathds R}^1\times (0, 2{\rm M}))$, each, in the
form of warped product, equipped, in the Schwarzschild spherical
co-ordinate system $x=p \times s$, where $p=(t, r)$ on $P=P_{\rm I}
\cup P_{\rm II}$ and $s=(\vartheta, \varphi)$ on $S^2$, with the
metric $g=g_{\rm P} + r^2 g_{\rm S}$, where $g_{\rm P}=- h(r)dt\otimes
dt+ {dr \otimes dr}/{h(r)}$, $h(r)=1 - {2{\rm M}}/{r}$ ($M>0$), and 
$g_{\rm S}$ the standard metric on $S^2$. $D$ is the Levi-Civita
connection of $g$. 

It is also quoted here, for notational purposes, the very basics of
the Kruskal space-time that are referred to in the following
section. The Kruskal plane $Q$ is obtained by joining the two
connected components $P_{\rm I}$ and $P_{\rm II}$ via a diffeomorphism
$f: {\mathds R}^+ \longrightarrow (-{2 {\rm M}}/{e}, +\infty)$ by
$f(r)=(r-2{\rm M})e^{\frac{r}{2{\rm M}}-1}$, ${\rm M}>0$. In terms of
the Kruskal null co-ordinate functions $k=(u,v) \in {\mathfrak
  F}({\mathds R}^2, {\mathds R})$ on the $u-v$ plane,
$Q=k^{-1}\Big(\{(u,v) \in {\mathds R}^2: uv> -{2 {\rm
    M}}/{e}\}\Big)$. The Schwarzschild co-ordinate 
functions $p=(t,r)$ on $Q$ become $t(u,v)=2{\rm M} {\rm log} |{v}/{u}
|$ on $Q-Q_{\rm H}$, where $Q_{\rm H}=k^{-1}(\{(u,v) \in k(Q):
uv=0\})$, and $r(u,v)=f^{-1}(uv)$ on $Q$. Thus, the metric on $Q$
reads $g_{\rm Q}=F(r) (du\otimes dv + dv \otimes du)$, where
$F(r)=\frac{8{\rm M}^2}{r}e^{1-\frac{r}{2 {\rm M}}}$, $\forall r \in 
{\mathds R}^+$. The Kruskal space-time, which is inextendible and
incomplete, is the warped product $K=Q\times_{\rm r} S^2$, equipped with the
metric $g_{\rm K}=g_{\rm Q} + r^2 g_{\rm S}$. The four connected components,
$\{Q_{j}\}_{j={\rm I}}^{j={\rm IV}}$, of  $Q-Q_{\rm H}$ consisting of four
open quadrants in the Kruskal plane, when lifted to $K$, yield $K_{\rm III}
\approx K_{\rm I} \approx N$ and $K_{\rm IV} \approx K_{\rm II} \approx B$,
by virtue of the isometries that preserve the Schwarzschild functions $p$.
It is therefore sufficient to consider half of the Kruskal space-time.

The notions of a family of observers and an observer field (aka a reference
frame) are used interchangeably throughout, the definition of which is
adopted from Ref.~\cite{SachsWu1977}.  Namely, an {\it observer field} $Q$
on a space-time $ {\cal M}$ is a vector field each of whose integral curves is an
observer,  where an {\it observer} is a future-directed  time-like curve
$\gamma: \mathscr E \longrightarrow  {\cal M}$ with unit speed.

\section{Measurement by non-geodesic observers}
\label{nongeodesic} 

\subsection{Static observer field}

The Schwarzschild exterior $N$, oriented by a nowhere-vanishing volume form
$\Omega$ and time-oriented by defining the time-like  Killing vector field
$\frac{\partial}{\partial t}$ to be future-directed, is static with respect
to the Schwarzschild observer field $U=\frac{1}{\sqrt{h(r)}}
\frac{\partial}{\partial t}$, where $h(r) \in (0, 1)$ on $N$. Since $D_{U}
U=\frac{{\rm M}}{r^2}\frac{\partial}{\partial r}$, $U$ is not geodesic. The
Schwarzschild  interior $B$, however, is not static with respect to any
observer field. Hence, Schwarzschild observers, i.e., integral curves of
$U$, exist only in $N$, as $U$ becomes space-like in $B$. What inhabits $B$,
nonetheless, is a geodesic observer field
$V=-\sqrt{-h(r)}\frac{\partial}{\partial r}$, which is proper-time
synchronizable since the dual frame field is $\theta^{\rm V}=-dg$, where
$g(r)=\int\limits_{r^\prime =r_0}^{r}\frac{dr^\prime}{\sqrt{-h(r^\prime)}}$,
for some $r_0 \in {\mathds R}^+$.

The Brown-York QLE measured by the Schwarzschild
observer field in $N$ is 
\begin{equation}
{\rm QLE}(r) = r [1 - \sqrt{h(r)}].
\label{eqn:qleN}
\end{equation}
It is related to the ADM energy ${\rm M}$ for a
spherically symmetric matter distribution via \cite{Brown:1992br}
\begin{equation}
{\rm M} = {\rm QLE}(r_*) - \frac{{\rm QLE}^2(r_*)}{2 r_*},
\label{eqn:ADM}
\end{equation}
where $r_*$ is the outer boundary of the distribution and
${\rm M}=m(r_*)$. The negative energy term on the right-hand side is
interpreted in \cite{Brown:1992br} as the negative energy outside of
the matter distribution that equals the Newtonian gravitational
binding energy associated with building a spherically symmetric shell
with matter-energy ${\rm QLE}$ and radius $r_*$. In what follows, a
hypothetical quasi-equilibrium procedure of constructing a spherically
symmetric shell of gravitational mass ${\rm M}$ is described. The negative
of the work done by the static Schwarzschild observer field $U$
to build up such a shell is explicitly computed.

Initially, no mass distribution is present. The Schwarzschild observer
field $U$ starts moving from infinity, in a precisely spherically
symmetric manner, a dust-shell of gravitational mass $dm$ and of zero
thickness to a fixed radius $r_*$ which is outside of the horizon
radius. It is assumed that each step is within quasi-equilibrium in
the sense that staticity of the space-time is not sabotaged. It
appears also legitimate to require that $dm$ be small enough so that
the influence of the dust shell on $N$ is negligible. Moreover, the
accumulation of mass  on the shell at radius $r_*$ does not incur an
expansion in its  radius. Hence, no mechanism is provided to sustain
the shell at rest; eventually, the shell of total gravitational mass
${\rm M}$ serves as an equivalent description of a spherically symmetric relativistic
star of total gravitational mass ${\rm M}$ and radius $r_*$ seen from
outside. More precisely, each of the instantaneous observers at $z \in
N$ with co-ordinates $x(z)=(t, r,\vartheta, \varphi)$, carrying in
$T_zN$ an orthonormal frame $X=\{U, X_2, X_3, X_4\}$, is referred to
as $(z, X)$. The comoving dust-shell flow is modeled by its energy
stress \cite{SachsWu1977}  $T=\sigma \theta^{\rm U} \otimes
\theta^{\rm U}$, where the gravitational mass surface density $\sigma$
measured by $(z, X)$ takes the form, in the Schwarzschild spherical
co-ordinate system $x$,  $\sigma \circ x=\frac{dm}{4 \pi r^2
  \sqrt{h(r, m)}} \delta(x^1-r)$ and $\theta^{\rm U}$ is the dual of
$U$. Here, a more explicit notation $h(r, m)=1-\frac{2 m}{r}$ is
employed purely to emphasize the dependence on the mass parameter. The
use of the Dirac measure in the co-ordinate representation of $\sigma$
is but to indicate that  ${\rm supp}(\sigma\circ x)$ is the shell of
radius $r$ with zero thickness.
 
Before carrying out the computations, it seems worthwhile clarifying
here the meaning of ``work''. The absence of a globally defined
inertial reference frame in $N$ prohibits any definition of work in
the Newtonian fashion. However, the comoving feature of the dust-shell
described above facilitates the evaluation of the work done by the
observer field $U$ in the general relativistic context. An intuitive
physical picture of the hypothetical shell-building process resembles
very much a bucket brigade, where, by virtue of the spherical
symmetry, the observer field $U$ is regarded as a continuum of
radially lined up observers each exercising a certain amount of work
moving a bit of the gravitational mass of the shell infinitesimally at
their respective location. The total work, being a scalar, is
registered as the sum of the infinitesimal work from each
step. Precise calculations are as follows.

Suppose there exists already a spherical shell of gravitational mass
$m$ at radius $r_*$. In the instantaneous 
inertial frame of each $(z, X)$, with $x^1(z)=r>r_*$, the measured
infinitesimal gravitational mass density $\sigma=T(U, U)$ gets moved
at the acceleration  $A=D_U U=\frac{m}{r^2 \sqrt{h(r, m)}} X_2$ by an
infinitesimal ``proper'' radial distance
$-\theta^2=-\frac{dr}{\sqrt{h(r, m)}}$. The infinitesimal work done in
each such step amounts, in the continuous limit, to the total work
done by the observer field $U$ throughout the entire moving process:
\begin{eqnarray}
&& W_U=\varint\limits_{0}^{{\rm M}} \varint\limits_{+\infty}^{r_*}
\varint\limits_{S^2} (\sigma\circ x)|_{x^1=r} g(A, X_2) \Omega_{S^2}
(-\theta^2) \cr && \qquad\qquad =-{\rm M}+r_*(1-\sqrt{h(r_*)}), \nonumber
\end{eqnarray}
where $\Omega_{S^2}$ is the standard volume form on $S^2$ and
$h(r_*)=1-\frac{2{\rm M}}{r_*} \in (0,1)$. Written in a more suggestive
form, the above relation, ${\rm M}={\rm QLE}(r_*)+ (-W_{\rm U})$, is
precisely Eq. (\ref{eqn:ADM}). 

Recall the known inequality \cite{Bizon:1990nk} on the total
gravitational energy of a spherically symmetric distribution of matter
which is instantaneously at rest: ${\rm M} \leq {\rm M}_{\rm p} - E_{\rm B}$,
where ${\rm M}$ and ${\rm M}_{\rm p}$ are the total gravitational and the proper
mass of the distribution, respectively, and $E_{\rm B}$ is the binding
energy. The equality is achieved if and only if the general
relativistic configuration possesses the least binding energy, i.e., a
spherically symmetric shell as in the Newtonian limit. When ${\rm
  QLE}(r_*)$ is understood to be the proper energy within the radius
$r_*$ as in \cite{Brown:1992br}, $W_{\rm U}$ gives the least binding
energy required for the spherically symmetric shell configuration. 
 
\begin{figure}[ht]
\includegraphics[scale=0.6]{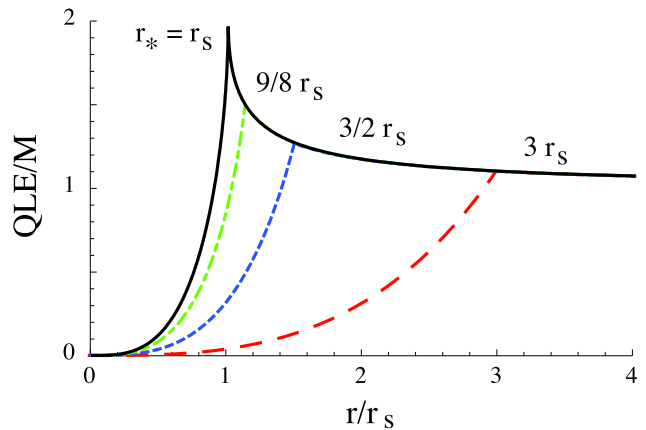}
\caption{The BY QLE for stationary, non-geodesic observers outside a constant 
desity star of
mass $M$ and radius $r_*$ is shown. Four cases are shown
corresponding to $r_*/r_S=1$ (unstable, solid), and
$r_*/r_S=9/8,\, 3/2,\, 3$ (long-short-dashed, short-dashed, long-dashed 
lines) where $r_S=2{\rm M}$ is the Schwarzschild radius of the star. In all
cases, 
the slope of the QLE is discontinuous at the surface of the star.}
\label{fig1} 
\end{figure}

The geometry of the exterior of a spherically symmetric relativistic star is
known to be modeled by the Schwarzschild exterior $(N, g)$, provided that
the horizon is frozen inside the surface of the star. As another example,
the QLE of a spherically symmetric star is easily computed. For a mass
density $\rho(r)$, the gravitational  mass contained within a radius $r$ is
$m(r) = 4\pi \int dr\, r^2 \rho(r)$. Eq. (\ref{eqn:qleN}) indicates
that the slope of the QLE is discontinuous across the surface of the star.
Furthermore, based on the positivity of the mass density, the QLE is a
maximum at the surface of the star. Specifically, consider the case
of a constant density star of radius $r_*$ and gravitational mass ${\rm M}
\in (0, \frac{4}{9}r_*)$, with Schwarzschild radius $r_S = 2 {\rm M} <
r_*$. The QLE is 
\begin{eqnarray}
{\rm QLE}&=& 2 {\rm M}\frac{r}{r_S} [1 - \sqrt{ h(r) } ]\cr
h(r) &=& 1 - \frac{r_S}{r}
\left(\left(\frac{r}{r_*}\right)^3\Theta(r_*-r) - \Theta(r-r_*)\right)  
\label{eqn:star}
\end{eqnarray}
where $\Theta$ is the Heaviside function. The QLE grows inside the
star from the center to the surface, whereupon it decays to ${\rm M}$ at infinity 
(see Figure~\ref{fig1}). The left (resp. right) radial derivative of the
QLE at $r_*$ tends to $+\infty$ (resp. $-\infty$) as $r_{\rm S}$
approaches $r_*$. 

\subsection{Non-static observer field}
 
It is natural to seek the QLE inside the horizon, in $B$ as well. The
investigation has been carried out recently in Ref.~\cite{Lundgren:2006fu},
wherein  an attempt has been made to continue the family of static,
non-geodesic Schwarzschild observers from $N$ into $B$. The result is that
the unit normal of $S^2$ as embedded in the space-like hypersurface $\Sigma$
is chosen to be time-like in $B$ (c.f. Eq.(16) in \cite{Lundgren:2006fu}),
thereby forcing the observer field to be space-like in $B$. This unphysical
choice for the observers seems incompatible with the construction of the
Brown-York QLE. The result in \cite{Lundgren:2006fu}, for
Schwarzschild space-time, is invoked in \cite{Blau:2007wj} to establish the
relationship between the Brown-York QLE measured by the
radial geodesic observers in $N \cup B$ and the effective potential of the
radial time-like geodesics. However, the observers used in
\cite{Lundgren:2006fu} are not geodesic and are valid in $N$ only, which
makes the analysis in \cite{Blau:2007wj} less justified. If instead
the geodesic observer field $V$ is employed in $B$, then the resulting
QLE is simply ${\rm QLE} = r$, which matches Eq. (\ref{eqn:qleN})
at the black hole horizon. 
Similar calculations in more general
settings have been carried out in \cite{Booth:1998eh}, where the
``non-orthogonal'' boundaries are taken into account. The
non-orthogonality comes about by flowing the 2-dimensional closed
orientable space-like surface $S$ along an observer field $T$ that
differs from the unit time-like vector field $U$ that is orthogonal to
the 3-dimensional space-like hypersurface into which $S$ is
isometrically embedded. Such scenarios are of physical interest when
the so called ``boosted'' observer fields are considered. As a
consequence, a generalized reference embedding scheme has to be
devised to adapt the foliation of $S$ along the flow of $T$. In
particular, for the Schwarzschild space-time $N \cup B$, a natural
choice of $U$ in $N$ is the Schwarzschild static observer field, with
respect to which the observer field $T$ can be put in the form such as
Eq.(30) in Ref.~\cite{Booth:1998eh}. The different time orientation in
$B$, however, spoils the foliation by the privileged static observer
field and thus invalidates the choice of $T$ for $N$. Additional
comparison with the results in Ref.~\cite{Booth:1998eh} for  the
geodesic observer field in Schwarzschild geometry is given in
Sec.~\ref{geodesic}. In addition, so far as measurements are
concerned, it is always possible to associate to the observer field
2-dimensional closed orientable and space-like surfaces in the
orthogonal fashion as shown in what follows.

The crux of the matter is an observer field that can penetrate the
Schwarzschild horizon $r_S=2{\rm M}$. Such observer fields do exist. It is
perhaps more convenient to use the Kruskal space-time $K$. Note that the
radius function $r$ is well defined on all of $K$, whereas the time function
$t$ is defined only on $K-H$, where $H$ is the lift of $Q_{\rm H}$ and
represents the event horizon. By construction, $Y=\frac{1}{4 {\rm M}}
(v\frac{\partial}{\partial v}-u\frac{\partial}{\partial u})$ is a Killing
vector field on $K$ and equals $\frac{\partial}{\partial t}$ on $K-H$, but
$\frac{\partial}{\partial t}$ can be uniquely extended as a Killing field
over all of $K$. Hence, the Kruskal space-time can be time-oriented, in
accord with the physical interpretations on $N$, by requiring that
$\frac{\partial}{\partial t}$ be future-directed on $K_{\rm I}$. 

Clearly, the observer field of interest here cannot be $Y$, which changes
its chronological signature in $K_{\rm II}$. It is, however, chosen, among
others, to be $X_1=\frac{1}{\sqrt{2 F(r)}} (-\frac{\partial}{\partial
u}+\frac{\partial}{\partial v})$, with its dual
$\theta^1=\sqrt{\frac{F(r)}{2}} d\xi$, where $\xi=-u+v$. Hence, $X_1$ is
synchronizable, i.e., there exists a unique space-like hypersurface
$\Sigma$ of $ {\cal M}$, the distribution on which is annihilated by
$\theta^1$, and thus completely integrable. Furthermore, the
co-ordinate function $\xi \circ \iota= \xi_0 ({\rm const.})$, where
$\iota: \Sigma \hookrightarrow  {\cal M}$. Given $X_1$, there exists
$X_2=\frac{1}{\sqrt{2 F(r)}} (\frac{\partial}{\partial
  u}+\frac{\partial}{\partial v})$, with its dual
$\theta^1=\sqrt{\frac{F(r)}{2}} d\rho$, where $\rho=u+v$, such that
$X_2$ is space-like and $\theta^1(X_2)=0$. Consequently, $X_2 \circ
\iota =X_2$ and in $\Sigma$, there exists a unique space-like
hypersurface $S$ of $\Sigma$, the distribution on which is annihilated
by $\theta^2$ and thus completely integrable. Similarly, the
co-ordinate function $\rho \circ \eta = \rho_0 ({\rm const.})$, where
$\eta: S \hookrightarrow \Sigma$. Thus, on $S$, $\xi=\xi_0$ and
$\rho=\rho_0$, and correspondingly, $t=t(\xi_0, \rho_0)={\rm const.}$ and
$r=r(\xi_0,\rho_0)={\rm const.}$, which implies that $(\iota \circ
\eta)^\ast g=r^2 g_{\rm S}$. That is, $S=S^2$. To complete the frame field
for the observer field, pick the usual orthonormal frame on $S^2$, namely,
$X_3=\frac{1}{r}\frac{\partial}{\partial \vartheta}$ and $X_4=\frac{1}{r
{\rm sin \vartheta}} \frac{\partial}{\partial \varphi}$, with their duals
$\theta^3=r d\vartheta$ and $\theta^4=r {\rm sin \vartheta} d\varphi$,
respectively.

The core of the calculation in Brown-York QLE is the mean
curvature of $S^2$ as embedded in $\Sigma$. However, substantial
simplifications are readily available in the current setting. Since $X_2 \in
\Gamma(\Sigma, T\Sigma)$, the induced connection ${\widetilde D}$ of $D$ on
$\Sigma$ coincides with the Levi-Civita connection $D^{\Sigma}$ of $\Sigma$,
thus, the second fundamental form of $S^2$ as embedded in $\Sigma$ with its
normal $X_2$ is given by
\begin{eqnarray}
\II(X_l, X_l)&=&-\iota^\ast
g({\widetilde D}_{\eta_\ast X_l} X_2 \circ \iota, \eta_\ast X_l) \cr
&=& -g(D_{\eta_\ast X_l} \iota_\ast (X_2 \circ \iota), \eta_\ast X_l) \cr
&=&
\theta^2(D_{\eta_\ast X_l}\eta_\ast X_l), \nonumber
\end{eqnarray}
for $l=3, 4$. Standard Cartan calculus shows that $\II(\theta^3,
X_3)=\II(\theta^4, X_4)=\omega^2_3(\eta_\ast X_3)= -\frac{\sqrt{2 F(r)}}{8 M
r}(u+v)$, where $\omega$ is the connection 1-form on $M$. Hence, the trace
of the mean curvature is $k={\rm tr}\II=-\frac{\sqrt{2 F(r)}}{4 {\rm M} r}
(u+v)$. On the other hand, the flat-space reference embedding $\eta^0: S^2
\hookrightarrow {\mathds R}^3$ yields $k^0={\rm tr} \II^0=-\frac{2}{r}$.
Therefore, the Brown-York QLE is given by
\begin{eqnarray}
{\rm QLE}(t, r)&=&\frac{1}{8 \pi} \varint\limits_{S^2} (k-k^0)\Omega \cr
&=& r\Big[1-\frac{\sqrt{2 F(r)}}{8 {\rm M}}(u+v)\Big] \cr
&=& r\Big[1 - \delta \sqrt{h(r)} {\rm cosh} (\frac{t}{4 {\rm M}})\Big],
\nonumber
\end{eqnarray}
where $\Omega$ is the volume form of $S^2$ and $\delta= 1$ on 
$K_{\rm I}$ (exterior) and $-1$ on $K_{\rm II}$ (interior).

\begin{figure}[ht]
\includegraphics[scale=0.6]{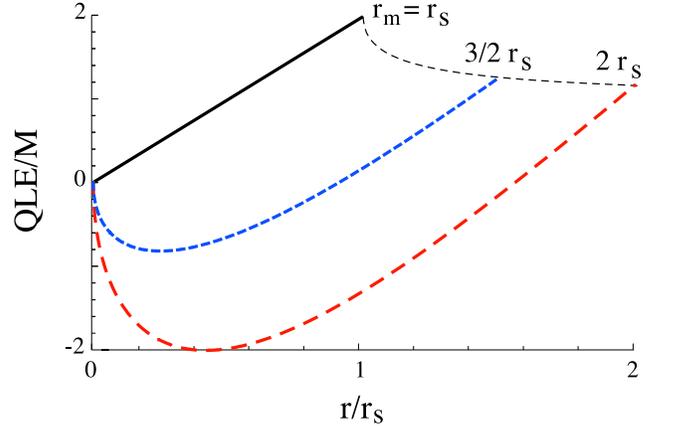}
\caption{The Brown-York QLE for non-geodesic observers, $X_1$, outside a
black hole  of mass ${\rm M}$ is shown. Three cases are shown, corresponding to 
maximum radius given by $r_{\rm max}/r_S =1, 3/2, 2$ (solid, short-dashed,
long-dashed lines). Note that QLE decreases smoothly as the observer crosses
the horizon.}
\label{fig3}  
\end{figure}

The QLE measured by the observers $X_1$ on $S^2$ depends on
both Schwarzschild functions $t$ and $r$, subject to the conditions
$\xi\circ\iota={\rm const}.$ and $\rho\circ\eta={\rm const}.$. This result
differs from that measured by the Schwarzschild observers $U$ in $N$ because
$X_1$ and $U$ produce different foliations of the space-time manifold. The
key difference, though, is that $X_1$ is an observer field over all of $K$,
while $U$ is defined on $N \approx K_{\rm I}$ only.  
The QLE is illustrated in Fig.~\ref{fig3}. In contrast to the case of
the static observer field, $D_{X_1}X_1=-\frac{1}{16{\rm
M}}(u+v)(\frac{1}{r(u,v)}+ \frac{1}{2{\rm M}})\sqrt{2 F}X_2$, the
work done by $X_1$ to build up the spherically symmetric shell of
gravitating mass ${\rm M}$ and radius $r_*$ is negative for $r_* \in {\rm
QLE}^{-1}({\mathds R}^-)$.

\section{Measurement by geodesic observers}
\label{geodesic}

A different type of observer field of physical interest is the geodesic
observer field in $N \cup B$, which crosses the event horizon. The
isometries of $S^2$ admit a system of the Schwarzschild spherical
co-ordinates with respect to which each of the geodesic observers, $\gamma:
{\mathscr E} \longrightarrow  {\cal M}$, is initially equatorial, i.e., $\vartheta
\circ \gamma \equiv \frac{\pi}{2}$, and thus, $X_1= {\dot
\gamma}=\frac{E}{h(r)}\frac{\partial}{\partial t} + \delta \sqrt{E^2 -
V(r)}\frac{\partial}{\partial r} + \frac{L}{r^2}\frac{\partial}{\partial
\varphi}$, depending upon whether the geodesic is ingoing ($\delta=-1$) or
outgoing ($\delta=+1$). Here the constants $L$, the angular momentum per
unit mass of the observer, and $E$, the energy per unit mass at infinity,
are related via the energy equation $E^2 =(\frac{d}{d\tau} (r \circ
\gamma))^2 + V(r)$, with $V(r)=(1+\frac{L^2}{r^2}) h(r)$ being the effective
potential. Of course, $X_1$, as a unit vector, is not defined on the horizon
$H$, in the Schwarzschild spherical co-ordinates, as usual; the energy
equation, nonetheless, holds by continuity.

For simplicity, only radial geodesics ($L=0$) are considered, in which case
two kinds of ordinary orbits are available: (1) the crash orbit ($E^2 \in
(0,1)$), where ingoing observers crash directly into the singularity at
$r=0$ and outgoing observers shoot out to a turning point $r_{\rm max}$
($E^2=h(r_{\rm max})$) and then back into crash; (2) the crash/escape orbit
($E^2 \in [1, +\infty)$), where ingoing observers crash while outgoing
observers escape to infinity. In $N \cup B$, the radial geodesic observer
field $X_1=\frac{\partial}{\partial \xi}$ is proper time synchronizable
since the dual frame $\theta^1=d\xi$, where
$\xi=Et-\int\limits_{r^\prime=r_0}^{r} \delta \frac{\sqrt{E^2 -
h(r^\prime)}}{h(r^\prime)}dr^\prime$. The rest of the construction follows
the same lines as in Sec.~\ref{nongeodesic}. There exists a space-like
$X_2=\frac{1}{\sqrt{E^2-h(r)}}\frac{\partial}{\partial \rho}$ chosen to be
consistent with the positive orientation of $\Sigma$ as embedded in $M$.
$X_2$ is synchronizable since $\theta^2=\sqrt{E^2-h(r)} d\rho$, where
$\rho=-\delta t + \int\limits_{r^\prime =r_0}^{r} \frac{E
dr^\prime}{h(r^\prime) \sqrt{E^2 - h(r^\prime)}}$. Hence, there exists a
co-dimension 1 embedding $\eta: S=S^2 \hookrightarrow \Sigma$ such that
$\rho \circ \eta=\rho_0 ({\rm const.})$. The mean curvature of $S^2$ as
embedded in $\Sigma$ with the unit normal $X_2=X_2 \circ \iota$ is given by
$k={\rm tr}\II=- 2 \frac{E}{r}$, where $r=r(\xi_0, \rho_0)$, with the
reference part intact. Therefore, the Brown-York QLE measured
by the geodesic observer field in $N \cup B$, for crashing orbits, is ${\rm
QLE}(r)=r(1-E)$, where $r \in (0, r_{\rm max}]$. Note that $E(r)$ remains
valid at $r=2{\rm M}$ by continuity.

\begin{figure}[ht]
\includegraphics[scale=0.6]{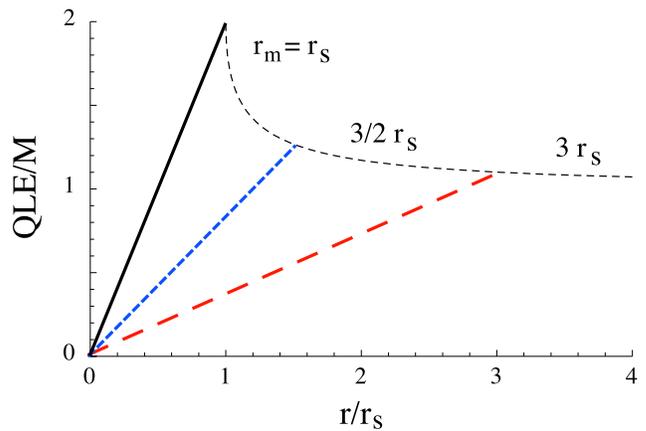}
\caption{The Brown-York QLE for geodesic observers outside a black hole of
mass $M$ is shown. Three cases are shown, corresponding to bound
trajectories with energy per unit mass $E = \sqrt{h(r_{\rm max})}$ starting
at a maximum radius given by $r_{\rm max}/r_S =1, 3/2, 3$ (solid,
short-dashed, long-dashed lines). Note that QLE decreases smoothly as the
observer crosses the horizon. In the case of an observer starting at $r_{\rm
max}/r_S \to \infty$, the QLE is zero.} 
\label{fig2}  
\end{figure}

In contrast to the analysis in \cite{Blau:2007wj}, the QLE
measured by the geodesic observer field does not seem to be related to the
effective potential in a non-trivial fashion. It is, however, simply linear
in the radius function $r$. At the turning point $r_{\rm max}$ for crash
orbits, ${\rm QLE}(r_{\rm max})=\frac{2{\rm M}}{1+E}$. At the event horizon,
${\rm QLE}(r)$ remains valid by continuity, and  ${\rm QLE}(r=2{\rm
M})=2{\rm M} (1-E)$.  A critical  situation is when $E^2=1$ and each
observer in this geodesic observer field starts at rest from infinity. In
the course of freely falling towards the singularity, the QLE
measured by this family of geodesic observers vanishes identically.
Interestingly, when $E^2>1$, the QLE become negative. This
result follows from the definition of the Brown-York QLE, whose sign is
determined by that of the sectional curvature of the 3-dimensional
space-like hypersurface $\Sigma$ into which the 2-dimensional consistently
oriented, closed, space-like surface $S$ is isometrically embedded. For the
geodesic observer field $X_1$, the sectional curvature of $\Sigma$ is
$K(\Sigma)=\frac{1-E^2}{r^2}$, which is positive ($E^2 \in (0,1)$), zero
($E^2=1$), or negative ($E^2 \in (1, +\infty)$). When compared to the
reference term evaluated in ${\mathds R}^3$, the Brown-York QLE becomes
positive, zero, or negative, respectively. In Figure~\ref{fig2}, different
QLE measurements are plotted for illustrative purposes. 

These results are different from those measured by the radially infalling
observers with $E=1$ considered in Ref.~\cite{Booth:1998eh}. As explained in
Sec.~\ref{nongeodesic}, the discrepancy arises from the generalized
embedding scheme  (c.f. Eq.(40)) in \cite{Booth:1998eh}, in accord with the
non-orthogonal boundaries. Should an $S^2$ that is orthogonal to $T$ be used
in \cite{Booth:1998eh}, the same QLE as given here follows immediately.  

It is curious to note that a star with mass density $\rho \propto r^{-2}$,
albeit unphysical due to the existence of a curvature singularity at the origin,
gives rise to a well-behaved QLE. In view of Eq.(\ref{eqn:star})
for observers $U$ in $N$, and observers $V$ in $B$,
${\rm QLE}= 2 {\rm M}\frac{r}{r_S}[1 - \sqrt{h(r)}]$ with $h(r)=1 -
\frac{r_S}{r_*}\Theta(r_*-r) - \frac{r_S}{r}\Theta(r-r_*)$. The
non-geodesic, stationary observer QLE curves in the limiting case $r_*/r_S\to 1$
and $r_*/r_S= 3/2,\, 3$ coincide with the curves for geodesic observers shown in Figure~\ref{fig2}. 

The rest of this section is devoted to a discussion of the measurement
of the quasi-local momentum by a family of geodesic observers. Similar
analysis can be carried out, though more complicated, for non-geodesic
observers, as well.

Consider an orthonormal frame field $\{X_i\}_{i=1}^{i=4}$, in which $X_1$ is
time-like geodesic and (locally) synchronizable. $\forall p \in {\cal M}$, $T_p
 {\cal M}={\mathds R}X_1 \oplus {\rm R}_p  {\cal M}$, where  ${\rm R}_p  {\cal M}={\rm
span}_{\mathds R}\{X_i\}_{i=2}^{i=4}$. The Brown-York quasi-local momentum
density ${\mathscr J} \in R^\ast  {\cal M}$ is defined by \cite{Brown:1992br}
${\mathscr J}(Y_p)=-\frac{1}{8\pi}\Big(\Theta \iota^\ast g(Y_p, X_2) -
\II^{X_1}(Y_p, X_2)\Big)$, $\forall (p, Y_p) \in R^\ast  {\cal M}$, where
$\Theta={\rm div} X_1$ is the expansion of the (local) flow of $X_1$ and
$\iota: \Sigma \hookrightarrow  {\cal M}$ is the (local) embedding induced by $X_1$.
When restricted to the 2-dimensional orientable closed space-like surface
$S$ embedded in $\Sigma$, whenever $\Sigma$ is completely integrable,
$\mathscr J$ is the quasi-local momentum surface density $j$, as originally
given in \cite{Brown:1992br}. It is clear from the definition that the
reference term for the quasi-local momentum vanishes identically when the
flat-space reference is employed. Hence, only the physical part is
considered in what follows. 

Given a one-parameter family of time-like geodesics  $x: {\mathscr E} \times
(-\delta, +\delta) \longrightarrow  {\cal M}$ around the base, $\gamma:{\mathscr E}
\longrightarrow  {\cal M}$ given by  $\gamma u=x(u,0)$, where $\gamma_\ast
\frac{d}{du}=X_1$, the Fermi-Walker connection \cite{SachsWu1977} coincides
with the induced connection $\gamma^\ast D$ over $\gamma$. For economy of
notation, there is no differentiation among various connections whenever the
context is clear. Recall that a  {\it neighbor} of the base is given by a
Jacobi field,  $J \in \Gamma({\mathscr E}, R {\cal M})$, over $\gamma$, as those
Jacobi fields that are tangent to $\gamma$ are of scant importance. It is
then convenient to introduce the following natural decomposition: $J=J^\bot
+ J^\top$, where $J^\bot =\theta^2(J)$ and $J^\top =J-J^\bot$. Accordingly,
the 3-relative quasi-local momentum density measured by a neighboring
observer with respect to the base is  ${\mathscr J}(J)={\mathscr J}(J^\bot)
+ j(J^\top)$. Thus, the Brown-York quasi-local momentum surface density
$j(J^\top)$ measured by the neighbor is the component that is tangent to
$S$, the orthogonal complement being the normal stretch ${\mathscr
J}(J^\bot)$.

The relative quasi-local momentum density in the family of geodesic
observers is obtained with the help of the Raychaudhuri equation for
time-like geodesics. Further simplifications emerge as the space-time of
interest is Ricci-flat. Since $X_1$ is (locally) synchronizable, hence
irrotational \cite{SachsWu1977}, the Raychaudhuri equation becomes
$\frac{d}{du}\Theta=-\frac{1}{3} \Theta^2 -{\rm tr}(\sigma^2)$, where 
$\sigma$ is the shear of $x$. Hence,  $\frac{d}{du}{\mathscr
J}(J)=-\frac{1}{8\pi}\Big[ \Theta {\mathscr J}(J) - (\Theta^2 + \II^{X_1}
\cdot \II^{X_1}) \iota^\ast g (J, X_2) - g(R_{J X_1}X_1, X_2)\Big]$, where 
$\II^{X_1} \cdot \II^{X_1}=\sum\limits_{i=2}^{i=4} \sum\limits_{j=2}^{j=4}
\II^{X_1}(\theta^i,\theta^j)\II^{X_1}(X_i,X_j)$.  When restricted to $S$, it
yields
$$
8\pi \frac{d}{du} j(J^\top) + \Theta j(J^\top)=
g(R_{J^\top X_1}X_1, X_2),
$$ 
where $g(R_{J^\top X_1}X_1, X_2)$ is the tidal force exerted on the neighbor
tangent to $S$ in the $X_2$-direction.

It is hoped that the above analysis provides a slightly different
perspective towards understanding the dynamics of the quasi-local momemtum
measured by the family of geodesic observers. 

\vspace*{0.2cm}
\section{Summary and discussion}
\label{summary}

The measurements of the Brown-York QLE in Schwarzschild space-time by a
non-static, non-geodesic observer field that penetrate  the event horizon
and by a geodesic observer field have been computed. It has been shown that
due to a different space-time foliation induced by the observer field, the
measurement of QLE by the non-static, non-geodesic observer field depends on
both the radius function $r$ and the time function $t$. The QLE measured by
a geodesic observer field is, however, linear in the radius function $r$, and
can be positive, zero, or negative, depending upon the energy parameter of
the geodesic observer field. These results differ significantly from the
measurement by a static observer field in the Schwarzschild exterior $N$,
which is previously known. On the other hand, the Liu-Yau QLE is independent
of the choice of observer field in $N$ and coincides with the Brown-York QLE
measured by the static observer field in $N$. However, the Liu-Yau QLE is
not defined in Schwarzshild interior $B$ because the mean curvature vector
of the co-dimension 2 isometric embedding of $S$ into $M$ becomes
time-like. 

To explore more about the physical nature of QLE, a hypothetical process of
building a spherically symmetric massive shell by a static observer
field was considered. The gravitational mass of the spherically
symmetric shell is the QLE of the shell plus the negative
gravitational potential energy associated with the work exercised to
build the shell. The result holds for both the Brown-York and the
Liu-Yau QLE since they coincide in $N$ for the static observer
field. In addition, the QLE of a spherically symmetric star with
constant density in the interior is also calculated. The QLE grows
monotonically with respect to radial distance from the origin in the
interior and drops off in the exterior towards the ADM mass ${\rm M}$ at
infinity, with a cusp on the surface of the star. It seems to provide
certain physical justification for QLE as a viable measure of energy.  
 
For a geodesic observer field in the Schwarzshild geometry, a dynamic
relation of the Brown-York quasi-local momentum density has been noted. It is
valid for any Ricci-flat space-time. When a family of geodesic observers are
assigned on $S$ to carry out physical measurements, this relation is
expected to describe their relative dynamics. Generalization to more
realistic Fermi-Walker observer fields is straightforward but more
complicated. 

\begin{acknowledgments}
PPY wishes to thank Professor C. Sutton for useful discussions. 
RRC was supported in part by NSF AST-0349213.
\end{acknowledgments}
 


\end{document}